\documentclass[twocolumn,nofootinbib]{revtex4-1}
\usepackage{latexsym,epsfig,amssymb, amsmath,nicefrac}

              \newcommand{\rf}[1]{(\ref{#1})}

\def\bfone{\relax{\rm 1\kern-.35em 1}}

\newcommand{\be}{\begin{equation}}
\newcommand{\ee}{\end{equation}}
\newcommand{\ben}{\begin{displaymath}}
\newcommand{\een}{\end{displaymath}}
\newcommand{\bea}{\begin{eqnarray}}
\newcommand{\eea}{\end{eqnarray}}

\newcommand{\bean}{\begin{eqnarray*}}
\newcommand{\eean}{\end{eqnarray*}}

\newcommand{\vp}{\varphi}

\begin{document}



\title{{\Large The Unity of Cosmological Attractors}}

\author{Mario Galante${}^1$, Renata Kallosh${}^2$, Andrei Linde${}^2$ and Diederik Roest${}^1$}

\affiliation{{}$^1$Van Swinderen Institute for Particle Physics and Gravity, University of Groningen, \\ Nijenborgh 4, 9747 AG Groningen, The Netherlands, m.galante@rug.nl, d.roest@rug.nl}

\affiliation{{}$^2$Department of Physics and SITP, Stanford University, \\ 
Stanford, California 94305 USA, kallosh@stanford.edu, alinde@stanford.edu}



\begin{abstract}
Recently, several broad classes of inflationary models have been discovered whose cosmological predictions are stable with respect to significant modifications of the inflaton potential. Some classes of models are based on a non-minimal coupling  to gravity. These models, which we will call $\xi$-attractors, describe universal cosmological attractors (including Higgs inflation) and induced inflation models. Another class describes conformal attractors (including Starobinsky inflation and T-models) and their generalization to $\alpha$-attractors. The aim of this paper is to elucidate the common denominator of these models: their attractor properties stem from a pole of order two in the kinetic term of the inflaton field in the Einstein frame formulation, prior to switching to the canonical variables. We point out that $\alpha$- and universal attractors differ in the subleading corrections to the kinetic term. As a final step towards unification of $\xi$ and $\alpha$ attractors, we introduce a special class of $\xi$-attractors which is fully equivalent to $\alpha$-attractors with the identification $\alpha = 1+{1\over 6\xi}$. There is no theoretical lower bound on $r$ in this class of models.
\end{abstract}


\maketitle

\smallskip


\section{Introduction}

The data releases by WMAP and Planck attracted attention to a mysterious fact: Two different models, the Starobinsky model \cite{Starobinsky:1980te} and the Higgs inflation model \cite{Salopek:1988qh}, make the same prediction, well matching observational data - both of Planck2013 \cite{Planck} as well as Planck2014: In the leading approximation in $1/N$, where $N$ is the number of e-folds, the spectral index $n_s$ and tensor-to-scalar ratio $r$ are given by
 \be
  n_s  = 1-\frac{2}{N}\,, \qquad 
  r  =  \frac{12 }{N^2} \, .
\label{attr} \ee
This could be a coincidence, but the further investigation revealed the existence of several broad classes of different models having the same predictions in the leading approximation in $1/N$, practically independent of the details of the model. 

The first class of these theories were conformal attractors \cite{Kallosh:2013hoa}, which described a broad variety of different models including  the Starobinsky model.   Further investigation revealed the existence of $\alpha$-attractors \cite{Ferrara:2013rsa, Kallosh:2013yoa}, which generalized the models of conformal attractors, but predicted, for not too large values of the parameter $\alpha$, that  
 \be
  n_s  = 1-\frac{2}{N}\,, \qquad 
  r  =   \frac{12 \alpha }{N^2} \, .
\label{aattr} 
\ee
The Lagrangian of the $\alpha$-attractor models of a real scalar field $\phi$ looks as follows in Einstein frame:
\be
  \mathcal{L}_{\rm E} = \sqrt{-g} \left[ {1\over 2} R - \frac{ \alpha}{ (1- \phi^2 /6)^2}{ (\partial \phi)^2\over 2} -  \alpha f^2(\phi / \sqrt{6})  \right] \, . \label{Lag-alpha}
\ee
It was shown in \cite{Kallosh:2013hoa, Ferrara:2013rsa, Kallosh:2013yoa} that the predictions  \rf{aattr} of this class of models are stable with respect to major changes of the inflaton potential, which has a functional freedom in terms of an arbitrary $f$. In this context, the Starobinsky model \cite{Starobinsky:1980te} corresponds to a special choice for this function with $\alpha = 1$.

Note that both the kinetic and potential energies have an overall coefficients $\alpha$. While the former appears in all versions of $\alpha$-attractor models, the latter is a matter of choice since the functions $f$ are nearly arbitrary. However, by placing $\alpha$ in from of it one reaches an important goal: In these classes of theories,  the parameter $r$ is proportional to $\alpha$, but the parameter $n_{s}$ {\it and} the amplitude of scalar perturbations of metric are independent of it.

Another class of models \cite{Salopek:1988qh,Kallosh:2013tua,Giudice:2014toa} described cosmological attractors with a non-minimal coupling to gravity:
\be
{\cal L}_{\rm J}=\sqrt{-g} \left[ \tfrac{1}{2} \Omega (\phi) R - \tfrac{1}{2} K_{\rm J} (\phi) (\partial \phi )^2  -V_{\rm J}(\phi) \right] \, ,
\label{lagJ}
\ee  
which we refer to as Jordan frame. 
For $\Omega = 1 +\xi\phi^{2}$, $V_{\rm J} = \lambda\phi^{4}$ and $K_{\rm J} = 1$ it described the Higgs inflation \cite{Salopek:1988qh}. In a more general class of models one retains the same functional relation between the non-minimal coupling and scalar potential,
 \be
   V_{\rm J}(\phi) = \frac{\lambda}{\xi^2} (\Omega (\phi) - 1)^2 \,, \label{U-Omega}
\ee
but allows for a different form of these functions. For instance, the universal attractor models are based on $\Omega = 1 +\xi f(\phi)$, where the function $f(\phi)$ can be arbitrary, and $K_{\rm J} = 1$ \cite{Kallosh:2013tua}. 

In the class of induced inflation models \cite{Giudice:2014toa} one has $\Omega = \xi f_{\rm ind}(\phi)>0$ and $K_{\rm J} = 1$. This class of theories is equivalent to the class of universal attractors up to the redefinition $f_{\rm ind}(\phi)  = f(\phi) +\xi^{{-1}}$ \cite{Kallosh:2014rha}. However, it is convenient to consider these two classes of models separately, by defining universal attractors as the theories where $\Omega = 1$ in the limit $\phi\to 0$, and induced inflation as the theories where $\Omega =0$ in the limit $\phi\to 0$. (Induced inflation originally was introduced in \cite{Giudice:2014toa} to provide a better description of the Higgs inflation, but in fact the mass of the inflaton field in this scenario typically is many orders of magnitude higher than the Higgs mass.) The inflationary predictions of all of these models depend on $\xi$ but coincide with \rf{attr} in the large $\xi$ limit, and are stable with respect to certain further modifications of $V_{\rm J}(\phi)$ to be discussed in this paper. Other choices of $\Omega$ and $K_{\rm J}$ have been also discussed in the literature. In this paper, we will call all models of this type $\xi$-attractors.

In addition to models with one attractor point, there were double attractors  \cite{Kallosh:2014rga}; their predictions interpolated between the predictions of $\alpha$-attractors with small $\alpha$, or induced inflation at large $\xi$, and the predictions $r = 4(1- n_s)  = 8/N$ of the simplest chaotic inflation model ${1\over 2} m^{2} \phi^{2}$  in the opposite parameter limit.

Despite the deepening understanding of the nature of these models \cite{Kallosh:2014rha}, a direct link between the models with non-minimal coupling and the $\alpha$-attractors was missing, and their predictions coincided with each other only in certain limits. In this paper we aim to clarify both the relations and differences between these models. We will emphasize that {\it the robust inflationary predictions \eqref{aattr} are a consequence of the properties of the leading pole in the Laurent expansion of the kinetic term of the Einstein frame of any attractor} (see e.g.~\eqref{Lag-alpha} for the $\alpha$-attractors): the order of the pole determines $n_s$ while its residue fixes $r$. For $\xi$-attractors with non-minimal coupling $\Omega$, we will demonstrate that the Einstein frame kinetic term has an identical pole in the variable $1/\Omega$. It is this common denominator in the kinetic term that underlies the attractor properties of these inflationary models.

As an application of our framework, we will introduce a special class of $\xi$-attractor models with a well-chosen designer kinetic term $K_{\rm J}$. For $\xi > 0$, they have the same observational predictions as $\xi$-attractors in the large $\xi$ limit; in fact, for all positive $\xi$ they are equivalent to induced inflation models with $\Omega = \xi \phi^{2}$ and $K_{\rm J}=1$. However, we will demonstrate that these models are well defined both for $\xi > 0$ as well as for $\xi < -1/6$. It turns out that {\it this class of $\xi$-attractors is 
fully equivalent to  $\alpha$-attractors for}
 \begin{align} \alpha = 1+{1\over 6\xi}  \,,\label{alpha-xi}
 \end{align}
with $\xi > 0$ and well as with $\xi < -1/6$. We will call these models {\it special attractors.} This provides a unification of a broad class of different attractor models, schematically represented in Fig. 1.
\begin{figure}[htb]
\vspace*{-1mm}
\begin{center}
\includegraphics[width=9cm]{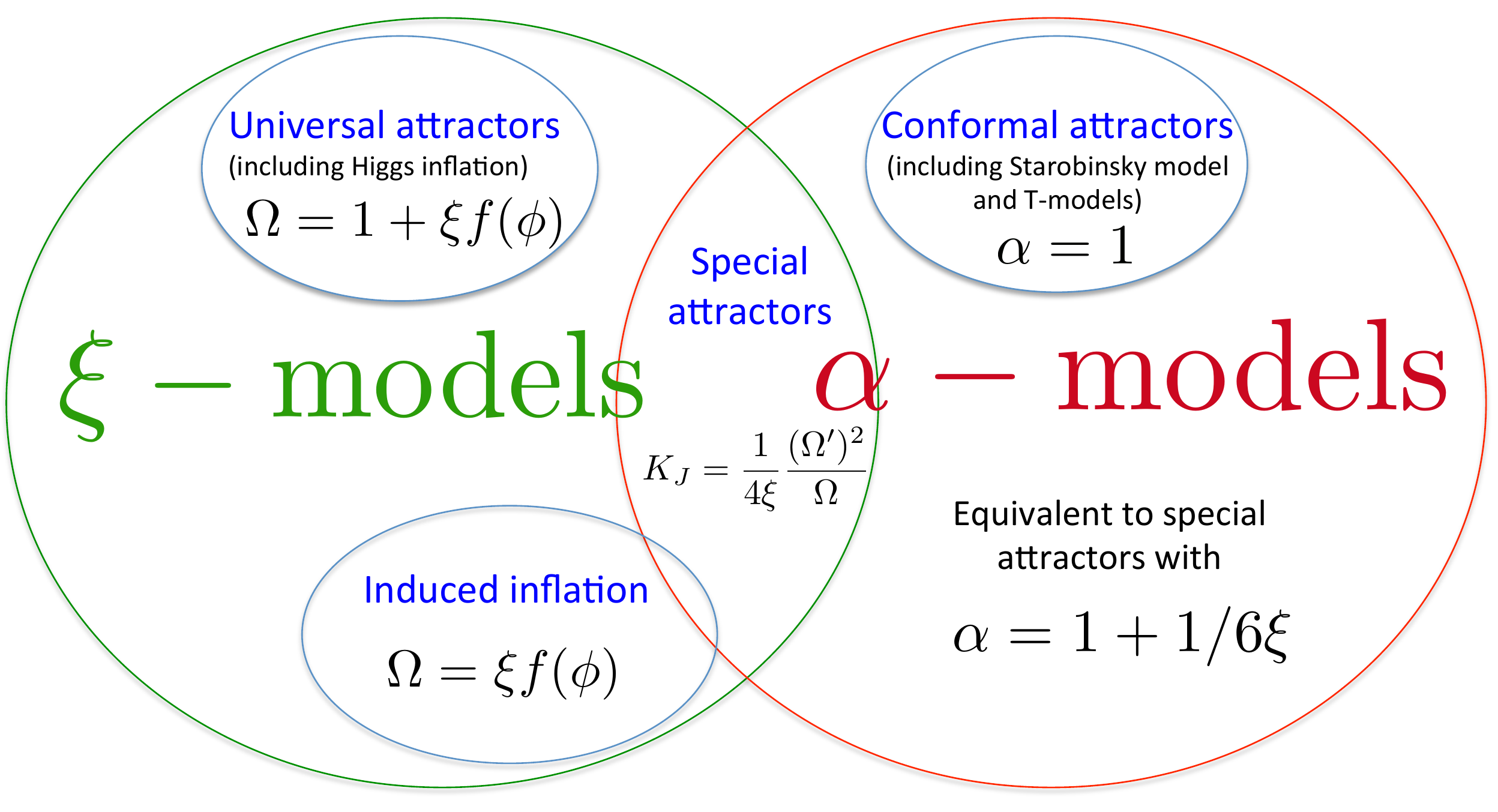}
\caption{\small Unification of cosmological attractors. The new class of $\xi$-attractors, which we called  {\it special attractors},  defined by \rf{lagEa}, are fully equivalent to $\alpha$-attractors with $\alpha = 1+{1\over 6\xi}$. }\label{uni}
\end{center}
\vspace{-0.7cm}
\end{figure}

The organization of this paper is as follows. We will start by emphasizing the role of poles in the kinetic formulation and demonstrate the relation to $\alpha$-attractors in section 2. Next, we move to non-minimal coupling and $\xi$-attractors in section 3, where we emphasize the relations between Jordan and Einstein frames, introduce the class of special attractors, and demonstrate the relations to induced inflation and the universal attractor. We conclude in section 4.

\section{Kinetic poles: $\alpha$-attractors}

\subsection{Kinetic formulation}

We will start by substantiating the claim in the introduction that the attractor nature of these theories stems from the property of any leading pole in the kinetic term. This simple observation can be phrased as
 \begin{quote} \it
  The inflationary predictions of models whose kinetic term is given by a Laurent series are determined by the order and the residue of the leading pole of the series.
 \end{quote}
In the above we have assumed minimal coupling to gravity, i.e.~Einstein frame, as well as a smooth scalar potential at the location of the pole. Such a model can be summarised as
  \be
  {\cal L}=\sqrt{-g} \left[ \tfrac{1}{2} R -  \tfrac{1}{2} K_{\rm E} (\rho)  (\partial \rho )^2 - V_{\rm E}(\rho) \right] \,.
 \label{kinetic}
\ee 
The case where $K_{\rm E}$ is given by a Laurent series (where we have assumed the pole to be located at $\rho=0$ without loss of generality) 
 \be
   K_{\rm E}= \frac{a_{p}}{\rho^p} + \ldots  \,, \qquad V_{\rm E} = V_0 ( 1 + c \rho + \ldots ) \,, \label{Laurent}
 \ee
is particularly interesting: it corresponds to a fixed point of the inflationary trajectory, which is characterised almost completely by the properties of this point. Indeed, in the limit of a large number of e-folds, one can assume that only the leading pole in $K_{\rm E}$ is relevant. This leads to the simple relation (where we will assume $p>1$ for simplicity)
\begin{equation}\label{N}
N=\int \frac{a_p}{c \rho^p} d\rho \sim \frac{a_p \rho^{1-p}}{c(p-1)} \,.
\end{equation}
Upon inverting this relation, one can calculate the spectral index and tensor-to-scalar ratio at leading order in $1/N$:
 \be
   n_s = 1 - \frac{p}{p-1} \frac1N \,, \qquad
  r=\frac{8 c ^{\frac{p-2}{p-1}} a_p^{\frac{1}{p-1}}}{(p-1)^{\frac{p}{p-1}}} \frac{1}{N^{\frac{p}{p-1}}} \,. \label{pole}
\ee
Indeed the spectral index depends solely on the order of the pole, while the tensor-to-scalar ratio also involves the residue. Note that this yields the same relation between the $1/N$ coefficient of the spectral index and the $1/N$ power of the tensor-to-scalar ratio as stressed in \cite{Mukhanov:2013tua}. Moreover, the kinetic formulation defines not only the power of $1/N$ but also the coefficient in the above formula for $r$. This is correct both for models with $p=2$, to be discussed later,  as well as for example hilltop inflation models \cite{Linde:1981mu} where $V_{\rm E} = V_0 (1- (\varphi/\mu)^n)$ with $p = 2 - 2/n$, where $n$ can be both negative and positive $n \geq 2$.

In what follows, we will be mainly interested in the case $p=2$: it is singled out as it allows for a superconformal and supergravity description, and arises as a consequence of a non-minimal coupling to gravity. In particular, we will show that all cosmological attractors can be brought to the form \rf{kinetic} with a kinetic term that has a pole or order two at a location where the scalar potential is perfectly smooth. In other words, all attractors have a common denominator in the Laurent expansion \rf{Laurent}. In this case, the general pole predictions \rf{pole} indeed lead to \rf{aattr} with the identification $ a_p = {3\over 2} \alpha$. (Note that for $p=2$ the constant $c$ drops out of \rf{pole}, as to be expected: it can be absorbed by a rescaling of the field $\rho$ while keeping the kinetic term invariant at leading order.) This provides a unified approach to their cosmological predictions, independent of the structure of the inflationary potentials - provided these are smooth at the point $\rho=0$.

\subsection{$\alpha$-attractors}

We will now demonstrate that the above is actually equivalent to the class of $\alpha$-attractors. To this end, we recall that the  original formulation of the theory of conformal attractors and $\alpha$-attractors \cite{Kallosh:2013hoa, Ferrara:2013rsa, Kallosh:2013yoa} was given in a non-canonical field $\phi$ as \eqref{Lag-alpha}. Its kinetic term has two poles of order two, related by symmetry $\phi \rightarrow - \phi$. Without loss of generality we will focus on the pole located at $\phi = \sqrt{6}$. Expanding around this pole, we find a Laurent expansion
 \begin{align}
   K_{\rm E} &  = \frac{3 \alpha}{2} \frac{1}{(\phi - \sqrt{6})^2} - \frac{\sqrt{6} \alpha}{4} \frac{1}{\phi - \sqrt{6}} + \ldots \,.
 \end{align}
Indeed we find the same leading pole of order two with residue $\tfrac32 \alpha$, in addition to subleading terms. Similarly, for a generic choice of the function $f$, the scalar potential is a Taylor series around the point $\phi = \sqrt{6}$. 

By means of field redefinitions one can change the form of the subleading terms, and trade certain subleading corrections to others. For instance, in this case one can redefine the field $\phi$ into a new variable $\rho$, such that the kinetic term becomes only a pole in $\rho$, without additional terms. This can be performed by
 \be
   \frac{\phi}{\sqrt{6}} = \frac{1 - \rho}{1+ \rho} \,.
 \ee
The Lagrangian of the $\alpha$-attractor models \rf{Lag-alpha} in the new variables $\rho$ has
\be
 K_{\rm E} = \frac{3 \alpha}{2} \frac{1}{\rho^2} \,, \quad V_{\rm E} = \alpha f^2 \Big ({1-\rho \over 1+\rho } \Big) \, . \label{lag-mho}
\ee  
Finally, one can go to a canonical field $\varphi$ with $K_{\rm E} =1$, where the scalar potential reads 
\be\label{EinstPot}
V_{\rm E} =  \alpha f^2 \bigl( \tanh {\varphi \over \sqrt{6 \alpha}}\bigr).
\ee
For $\alpha = 1$ and monomial functions $f$ they coincide with the T-models from the theory of conformal attractors \cite{Kallosh:2013hoa}.

Note that the kinetic terms blows up at $\phi = \sqrt{6}$ or $\rho = 0$. While the subleading corrections are different, both cases have the same leading term: this corresponds to a pole of order two with residue $3 \alpha / 2$. It is this singularity that is responsible for the stability of predictions of these theories \rf{aattr} with respect to strong deformations of the inflationary potential near the boundary of the moduli space at $\rho =0$. Subleading corrections in either the Laurent expansion of the kinetic term or the Taylor expansion of the potential term are irrelevant in the large-$N$ limit.

In terms of the canonical scalar field, this boundary is located at $\varphi$ close to infinity. For generic functions $f$, the scalar potential will asymptote to a plateau at infinity and will have an exponentially suppressed fall-off with leading term $e^{- \sqrt{2/3 \alpha}\, \varphi}$. It is this leading term that determines all inflationary properties at large $N$.

\section{Non-minimal coupling: $\xi$-attractors}

\subsection{Special attractors} 

We will now address the relation between the $\alpha$-attractors with a pole in the kinetic term and the $\xi$-attractors based on a non-minimal coupling between the gravitational and inflationary sector. Therefore we generalize our starting point to the Jordan frame \rf{lagJ}. By means of a conformal transformation for $\Omega > 0$, it can be brought to the Einstein frame with 
  \be
 K_{\rm E} = \left( \frac{K_{\rm J}}{\Omega}+\frac{3{\Omega^\prime}^2}{2\Omega^2}\right)  \,, \quad V_{\rm E} = \frac{V_{\rm J}(\phi)}{\Omega^2} \, .
\label{lagE}
\ee

So far, only models with $K_{\rm J} = 1$ have been considered, where the parameter $\xi$ was a part of the choice of the function $\Omega(\phi)$ in \rf{lagJ}. Now we will define a new class of theories, which we will call {\it special attractors}. They will be defined by the following choice of functions in \rf{lagJ}:
\be
K_{\rm J}= {1\over 4 \xi} \frac{(\Omega^\prime)^2}{\Omega} \,, \quad 
V_{\rm J}(\phi) =  {\Omega^2}\, U(\Omega)\ .
\label{lagEa}
\ee
Thus we absorbed the $\xi$ dependence into the factor $K_{\rm J}$. 
Then the theory \rf{lagJ} in the Einstein frame becomes
 \be
{\cal L}_E=\sqrt{-g} \left[ \frac{R}{2}  -\frac{3\alpha}{4} \left(\partial \Omega \over \Omega\right)^{2} -U(\Omega)  \right] \,,
\label{lagEaA}
\ee
where
\be
 \alpha \equiv  1+ {1\over 6 \xi }  \ .
\label{lagEaAAA}
\ee
In this theory $\Omega$ becomes the field variable. Its kinetic term is exactly of the form \rf{Laurent} with a pole of order two and no subleading corrections. However, physically this does not correspond to the same limit: while the $\alpha$ attractors derive their attractor predictions from the region close to $\rho=0$, inflation in the $\xi$-attractors takes place at $\Omega$ very large. Therefore it is natural to identify 
\begin{align}
 \rho(\phi) = \Omega^{-1} (\phi)  \,.
 \end{align} 
 Note that a pole of order two is exactly invariant under this redefinition and retains the same form. 

In order for the kinetic energy to be well-defined, one has to require that $\alpha$ is positive. There are three regions of the parameter $\xi$; the condition $\alpha> 0$ is satisfied in the first two of them:
 \begin{itemize}
 \item $\xi>0$, with $\alpha >1$, or
 \item $   - \infty < \xi < -{1\over 6}$ corresponding to $0<\alpha<1$, while
 \item Intermediate regions with $- 1/6 < \xi < 0$ lead to a wrong sign of the Einstein frame kinetic term.
 \end{itemize}
The limiting case with $\alpha=1$ can be reached either in the limit $\xi \rightarrow \infty$ or $\xi \rightarrow - \infty$, while $\alpha = 0$ is accessible via $\xi \rightarrow -1/6$ from below.

It is important to take stock of the situation at this point. In particular, one can allow $\xi$ to become negative (and $\alpha$ to become smaller than one) at a very specific price: the Jordan frame kinetic term \rf{lagEa} has the wrong sign. While this could seem dangerous, for $   - \infty < \xi < -{1\over 6}$ this danger is in fact fictitious as it does not lead to negative kinetic terms and instability in the Einstein frame.

This phenomenon is reminiscent of the Breitenlohner-Freedman bound in Anti-de Sitter space. In that case, an apparent instability due to a negative mass can be cured by the non-trivial geometry provided the mass satisfies the BF bound \cite{Breitenlohner}. In our case, an apparent instability due to a negative kinetic energy can be cured by the non-minimal coupling in Jordan frame, provided the coefficient $1/(4 \xi)$ of the negative term in \eqref{lagEa} is sufficiently small such that $\alpha$ is positive.

One can represent the theory \rf{lagEaA} in terms of a canonically normalized inflaton field $\varphi$ as follows:
\be
\Omega = e^{\sqrt{2\over 3\alpha}\vp} \,,
\ee
and
\be
{\cal L}_E=\sqrt{-g} \left[ \frac{R}{2}- \frac{1}{2} (\partial \vp )^2-  
U\big (e^{\sqrt{2\over 3 \alpha } \vp}\big ) \right]
\label{can}\ee
For the special choice $U(\Omega) = \alpha f^2 \Big ({1-\Omega \over 1+\Omega} \Big)$, this theory coincides with the class of $\alpha$-attractors defined in \rf{lag-mho}, \rf{EinstPot}, with $V_{\rm E} = \alpha f^2 \bigl( \tanh {\varphi \over \sqrt{6 \alpha}}\bigr)$. In particular, for the simplest choice $f(x) = c x$, where $c $ is some constant, one finds the $\alpha$-generalization of the simplest T-model potential  \cite{Kallosh:2013hoa,Kallosh:2013yoa}
\be
V = \alpha c^{2}  \tanh^{2} {\varphi \over \sqrt{6 \alpha}} \ ,
\ee
and for $f(x) = {cx\over 1+x}$, which is equivalent to the choice $V_{\rm J} = c^{2}(\Omega-1)^{2}$, one finds a generalization of the Starobinsky 
potential, called $\alpha-\beta$ model \cite{Ferrara:2013rsa}
\be\label{albet}
V = \alpha c^{2}  \left(1-e^{-\sqrt{2\over 3\alpha}\vp}\right)^{2} \ .
\ee
More general choices of potentials are possible, e.g. one can add to $U(\Omega)$ corrections 
 \begin{align}
\Delta U(\Omega) =   \sum_{i=2}^{\infty} c_i \Omega^{-i} =   \sum_{i=2}^{\infty} c_i \rho^{i} \ . \label{generalization}
  \end{align} 
This results  in the subleading corrections in $e^{\sqrt{2\over 3\alpha}\vp}$, which do not affect the inflationary predictions in the large-$N$ limit.

\subsection{Relation to induced inflation}

Induced inflation is defined by \rf{lagJ} with  $\Omega = \xi f(\phi)$ an the scalar potential given by the usual relation \eqref{U-Omega}. As we already mentioned, this theory is well defined (i.e.~describes gravity instead of antigravity) only for $\Omega > 0$. Without any loss of generality, one can define this class of theories by conditions $\xi > 0$, $f(\phi) >0$. 
Then, independent of the function $f(\phi)$, which in principle can be chosen arbitrary, the inflationary predictions of this model coincide with  \eqref{attr} in the limit of $\xi \rightarrow +\infty$ \cite{Giudice:2014toa}. Moreover, in the opposite limit $\xi \rightarrow 0$ the predictions approximate those of quadratic inflation, again independent of the functional choice \cite{Kallosh:2014rga}. 

Remarkably, for the special case $\Omega = \xi \phi^2$ and $\xi >0$ the induced inflation model in the Einstein frame is also represented by the special attractor action \rf{lagEaA}, \rf{lagEaAAA}.  In this model $V_{\rm J} = \frac{\lambda}{\xi^2} (\Omega - 1)^2$, and the Einstein frame potential for $\alpha = 1$ is given by \rf{albet} with $c^{2} = {\lambda\over\xi^{2}}$. This choice of $c^{2}$ here is not required, it was motivated by the desire to implement the Higgs inflation scenario  \cite{Giudice:2014toa}. But the potential \rf{albet} is different from the Higgs inflation potential anyway: It is not symmetric with respect to the change $\vp\to-\vp$ and it does not contain an important part of the potential at intermediate values of $\vp$ where the potential is quartic in $\vp$. However, it is important that it belongs to the class of the special attractors.
Moreover, it allows for the same generalization \eqref{generalization} of the scalar potential.

\subsection{Relation to universal attractors}

Finally, we wish to emphasize how the universal attractor models of \cite{Kallosh:2013tua} are related to $\alpha$-attractors  and spell out how they fit in the present framework. The universal attractor models considered in \cite{Kallosh:2013tua} are defined by the choice $K_{\rm J} = 1$ and $\Omega =1+\xi f(\phi)$ for an arbitrary function $f(\phi)$. 

In the limit when $\xi \rightarrow \infty$ the inflationary predictions of these models coincide with those of the induced models with $\Omega \approx \xi f(\phi)$, as well as those of special attractors and $\alpha$ attractors for $\alpha\approx 1$. In this limit there is no need to make a choice $f(\phi)=\phi^2$ (as we did in the case of an exact relation between $\alpha$-models and generalized induced inflation models above). In the limit $\xi \rightarrow \infty$, the first term in \rf{lagE} can be neglected and we find
  \begin{align}
 K_{\rm E} = \frac{3}{2}  \frac{1}{\rho^2} \,, \quad 
  V_{\rm E} =  \frac{\lambda}{\xi^2}( 1 - \rho  )^2 \,, 
\label{lagE2}
\end{align}
where we have replaced the non-minimal coupling $\Omega(\phi)$ (which can be chosen arbitrarily) by its inverse $\rho$. Here we see again that the pole structure at $\rho=0$ allows us to deform the potential and, instead of the function \eqref{U-Omega} consider any function with additional terms with higher powers of $\rho=e^{- \sqrt{2\over 3} \vp}$. 

Moreover, one can calculate the subleading corrections to the above kinetic term that arise for finite values of $\xi$. For instance, in the case of Higgs inflation with $f = \phi^2$, the full kinetic term for the field $\rho$ is given by
 \begin{align}
  K_{\rm E} = \frac{3}{2} \frac{1}{\rho^2} + \frac{1}{4 \xi} \frac{1}{(1- \rho) \rho^2} = \frac{3\alpha}{2} \frac{1}{\rho^2} + \frac{1}{4 \xi} \frac{1}{\rho} \ldots + \,.
  \end{align}
While this has the same leading pole, subleading corrections will be different. A particularly acute difference with respect to the case of induced inflation, discussed in the previous subsection, is that the kinetic term is not necessarily positive definite. In particular, inflation takes place close to $\rho =0$, while the Minkowski vacuum is located at $\rho = 1$. In the latter regime, the second term will always be dominant. Therefore Higgs inflation does not allow one to take $\xi$ negative even in the Einstein frame, in contrast to induced inflation: in addition to the condition $\alpha>0$ from the inflationary regime, one also requires $\xi>0$ from the cosmological era following inflation.

\section{Discussion}

Provided the kinetic term of the inflaton is given by a Laurent series, its inflationary predictions are to a large extent determined by the properties of the leading pole, and therefore robust to changes to the subleading terms, either in the kinetic or the potential energy. Such a pole of order two underlies the attractor properties of both $\alpha$- and $\xi$-attractors and therefore yields the inflationary predictions \eqref{aattr}. 

Next, we have explicitly demonstrated the unity of these two types of attractors, either based on non-trivial kinetic terms or on non-minimal couplings: when transforming $\xi$-attractors from Jordan to Einstein frame, one obtains $\alpha$-attractors and vice versa. Moreover, we have emphasized that there is a special type of attractors whose kinetic term consists only of a single pole: both the original $\alpha$-attractors of \cite{Kallosh:2013yoa} as well as induced inflation \cite{Giudice:2014toa} are of this form. 

The introduction of generalized $\xi$ attractors including the special attractors  \rf{lagEa} opens a simple way towards the unification of all presently known cosmological attractors, as illustrated in Figure 1. We have shown that the class of the special attractors is equivalent to $\alpha$ attractors with $\alpha = 1+{1\over 6\xi}> 0$.  This relation between both parameters, which is one of our main results, embodies the two viable ranges $\xi>0 $ and $\xi < -1/6$. In the Jordan frame, only the first of these has a positive kinetic term, corresponding to $\alpha \geq 1$. However, similar to the BF bound, the theory is well defined for both cases: It has positive kinetic term in the Einstein frame and it does not exhibit any instability. There is no theoretical lower bound on $ r  =   \frac{12 \alpha }{N^2}$ in this class of models.


\section*{Acknowledgements}

We acknowledge stimulating discussions with B. Broy, S. Cecotti, G. Giudice,  S. Ferrara, H. M. Lee, K. Pallis, A. Van Proeyen, B. Vercnocke and A. Westphal. RK and AL are supported by the SITP and by the NSF Grant PHY-1316699 and RK is also supported by the Templeton foundation grant `Quantum Gravity Frontiers,' and AL is supported by the Templeton foundation grant `Inflation, the Multiverse, and Holography.' MG is supported by an Erasmus Mundus PEACE project scholarship.


\begin{thebibliography}{99}

\bibitem{Starobinsky:1980te} 
  A.~A.~Starobinsky,
``A New Type of Isotropic Cosmological Models Without Singularity,''
  Phys.\ Lett.\ B {\bf 91}, 99 (1980).
  V.~F.~Mukhanov and G.~V.~Chibisov,
``Quantum Fluctuation and Nonsingular Universe. (In Russian),''
  JETP Lett.\  {\bf 33}, 532 (1981)
  [Pisma Zh.\ Eksp.\ Teor.\ Fiz.\  {\bf 33}, 549 (1981)].
  A.~A.~Starobinsky,
``The Perturbation Spectrum Evolving from a Nonsingular Initially De-Sitter Cosmology and the Microwave Background Anisotropy,''
  Sov.\ Astron.\ Lett.\  {\bf 9}, 302 (1983).
  B.~Whitt,
``Fourth Order Gravity as General Relativity Plus Matter,''
  Phys.\ Lett.\ B {\bf 145}, 176 (1984).
  L.~A.~Kofman, A.~D.~Linde and A.~A.~Starobinsky,
``Inflationary Universe Generated by the Combined Action of a Scalar Field and Gravitational Vacuum Polarization,''
  Phys.\ Lett.\ B {\bf 157}, 361 (1985).
  
\bibitem{Salopek:1988qh}
  D.~S.~Salopek, J.~R.~Bond and J.~M.~Bardeen,
``Designing density fluctuation spectra in inflation,''
  Phys.\ Rev.\  \textbf{D40}, 1753 (1989).
   F.~L.~Bezrukov and M.~Shaposhnikov, ``The Standard
Model Higgs boson as the inflaton,'' Phys.\ Lett.\ B {\bf 659}, 703
(2008) [arXiv:0710.3755 [hep-th]].

\bibitem{Planck}
  P.~A.~R.~Ade {\it et al.}  [Planck Collaboration],
``Planck 2013 results. XXII. Constraints on inflation,''
  Astron.\ Astrophys.\  {\bf 571}, A22 (2014)
  [arXiv:1303.5082 [astro-ph.CO]].

\bibitem{Kallosh:2013hoa} 
  R.~Kallosh and A.~Linde,
``Universality Class in Conformal Inflation,''
  JCAP {\bf 1307}, 002 (2013)
  [arXiv:1306.5220 [hep-th]].
R.~Kallosh and A.~Linde,
``Multi-field Conformal Cosmological Attractors,''
  JCAP {\bf 1312}, 006 (2013)
  [arXiv:1309.2015 [hep-th]].
  
\bibitem{Ferrara:2013rsa} 
  S.~Ferrara, R.~Kallosh, A.~Linde and M.~Porrati,
  ``Minimal Supergravity Models of Inflation,''
  Phys.\ Rev.\ D {\bf 88}, no. 8, 085038 (2013)
  [arXiv:1307.7696 [hep-th]].

\bibitem{Kallosh:2013yoa} 
  R.~Kallosh, A.~Linde and D.~Roest,
  ``Superconformal Inflationary $\alpha$-Attractors,''
  JHEP {\bf 1311}, 198 (2013)
  [arXiv:1311.0472 [hep-th]].
  
\bibitem{Kallosh:2013tua} 
  R.~Kallosh, A.~Linde and D.~Roest,
 ``A universal attractor for inflation at strong coupling,''
  Phys.\ Rev.\ Lett.\  {\bf 112}, 011303 (2014)
  [arXiv:1310.3950 [hep-th]].


  
\bibitem{Giudice:2014toa} 
  G.~F.~Giudice and H.~M.~Lee,
``Starobinsky-like inflation from induced gravity,''
  Phys.\ Lett.\ B {\bf 733}, 58 (2014)
  [arXiv:1402.2129 [hep-ph]];
    C.~Pallis,
   ``Linking Starobinsky-Type Inflation in no-Scale Supergravity to MSSM,''
   JCAP {\bf 1404}, 024 (2014)
   [arXiv:1312.3623 [hep-ph]];
   C.~Pallis,
   ``Induced-Gravity Inflation in no-Scale Supergravity and Beyond,''
   JCAP {\bf 1408}, 057 (2014)
   [arXiv:1403.5486 [hep-ph]];
  C.~Pallis,
   ``Reconciling Induced-Gravity Inflation in Supergravity With The  
Planck 2013 \& BICEP2 Results,''
   JCAP {\bf 1410}, no. 10, 058 (2014)
   [arXiv:1407.8522 [hep-ph]].
  
\bibitem{Kallosh:2014rha} 
  R.~Kallosh,
``More on Universal Superconformal Attractors,''
  Phys.\ Rev.\ D {\bf 89}, no. 8, 087703 (2014)
  [arXiv:1402.3286 [hep-th]].
 
  
\bibitem{Kallosh:2014rga} 
  R.~Kallosh, A.~Linde and D.~Roest,
``Large field inflation and double $\alpha$-attractors,''
  JHEP {\bf 1408}, 052 (2014)
  [arXiv:1405.3646 [hep-th]];
  R.~Kallosh, A.~Linde and D.~Roest,
``The double attractor behavior of induced inflation,''
  JHEP {\bf 1409}, 062 (2014)
  [arXiv:1407.4471 [hep-th]];
  B.~Mosk and J.~P.~van der Schaar,
  ``Chaotic inflation limits for non-minimal models with a Starobinsky attractor,''
  arXiv:1407.4686 [hep-th].



\bibitem{Mukhanov:2013tua} 
  V.~Mukhanov,
  ``Quantum Cosmological Perturbations: Predictions and Observations,''
  Eur.\ Phys.\ J.\ C {\bf 73}, 2486 (2013)
  [arXiv:1303.3925 [astro-ph.CO]];
  D.~Roest,
  ``Universality classes of inflation,''
  JCAP {\bf 1401} (2014) 01,  007
  [arXiv:1309.1285 [hep-th]];
  J.~Garcia-Bellido and D.~Roest,
  ``The large-N running of the spectral index of inflation,''
  arXiv:1402.2059 [astro-ph.CO].
  
\bibitem{Linde:1981mu} 
A.~D.~Linde,
 ``A New Inflationary Universe Scenario: A Possible Solution of the Horizon, Flatness, Homogeneity, Isotropy and Primordial Monopole Problems,''
  Phys.\ Lett.\ B {\bf 108}, 389 (1982).
   A.~D.~Linde,
``Primordial Inflation Without Primordial Monopoles,''
  Phys.\ Lett.\ B {\bf 132}, 317 (1983).
  L.~Boubekeur and D.~H.~Lyth,
  ``Hilltop inflation,''
  JCAP {\bf 0507} (2005) 010
  [hep-ph/0502047].

\bibitem{Breitenlohner} 
  P.~Breitenlohner and D.~Z.~Freedman,
  ``Positive Energy in anti-De Sitter Backgrounds and Gauged Extended Supergravity,''
  Phys.\ Lett.\ B {\bf 115}, 197 (1982).

  
 \end{thebibliography}
\end{document}